\documentclass[sigconf]{acmart}
\AtBeginDocument{%
  }

\copyrightyear{2026}
\acmYear{2026}
\setcopyright{cc}
\setcctype{by}
\acmConference[KDD '26]{Proceedings of the 32nd ACM SIGKDD Conference on Knowledge Discovery and Data Mining V.2}{August 09--13, 2026}{Jeju Island, Republic of Korea}
\acmBooktitle{Proceedings of the 32nd ACM SIGKDD Conference on Knowledge Discovery and Data Mining V.2 (KDD '26), August 09--13, 2026, Jeju Island, Republic of Korea}
\acmDOI{10.1145/3770855.3818122}
\acmISBN{979-8-4007-2259-2/2026/08}

\usepackage{xspace}
\usepackage{enumitem}
\usepackage{multirow}

\usepackage[most]{tcolorbox}
\usepackage{xcolor}

\definecolor{caseblue}{RGB}{45, 75, 85}    
\definecolor{casebg}{RGB}{245, 250, 250}   

\newtcolorbox{mycase}[1]{
    enhanced,
    boxrule=0.8pt,
    colback=casebg,
    colframe=caseblue,
    sharp corners,
    top=2pt,                
    bottom=2pt,             
    left=5pt,               
    right=5pt,              
    toptitle=0pt,           
    bottomtitle=0pt,        
    detach title, 
    before upper={\vspace{0.2cm}}, 
    fonttitle=\bfseries\sffamily\small,
    coltitle=white,
    attach boxed title to top left={yshift=-3mm, xshift=5mm},
    boxed title style={
        sharp corners, 
        size=small, 
        colback=caseblue, 
        colframe=caseblue,
    },
    title=#1
}

\newcommand{\ie}{\emph{i.e., }}

\newcommand{\etc}{\emph{etc.}}

\newcommand{\cf}{\emph{cf. }}

\newcommand{\za}[1]{{\color{black}{#1}}}

\newcommand{\hyz}[1]{\iffalse #1 \fi}
\newcommand{\ours}{SIDReasoner\xspace}

\begin{document}
\title{Reasoning over Semantic IDs Enhances Generative Recommendation}


\author{Yingzhi He}
\email{heyingzhi@u.nus.edu}
\affiliation{%
  \institution{National University of Singapore}
  \city{Singapore}
  \country{Singapore}
}

\author{Yan Sun}
\email{sthen@mail.ustc.edu.cn}
\affiliation{%
  \institution{University of Science and Technology of China}
  \city{Hefei}
  \country{China}
}

\author{Junfei Tan}
\email{sober_clever@mail.ustc.edu.cn}
\affiliation{%
  \institution{University of Science and Technology of China}
  \city{Hefei}
  \country{China}
}

\author{Yuxin Chen}
\email{yuxin.chen@u.nus.edu}
\affiliation{%
  \institution{National University of Singapore}
  \city{Singapore}
  \country{Singapore}
}

\author{Xiaoyu Kong}
\email{kongxy@mail.ustc.edu.com}
\affiliation{%
  \institution{University of Science and Technology of China}
  \city{Hefei}
  \country{China}
}

\author{Chunxu Shen}
\authornote{Corresponding Authors.}
\email{lineshen@tencent.com}
\affiliation{%
  \institution{Tencent Inc.}
  \city{Shenzhen}
  \country{China}
}

\author{Xiang Wang}
\email{xiangwang1223@gmail.com}
\affiliation{%
  \institution{University of Science and Technology of China}
  \city{Hefei}
  \country{China}
}

\author{An Zhang}
\authornotemark[1]
\email{an.zhang3.14@gmail.com}
\affiliation{%
  \institution{University of Science and Technology of China}
  \city{Hefei}
  \country{China}
}

\author{Tat-Seng Chua}
\email{dcscts@nus.edu.sg}
\affiliation{%
  \institution{National University of Singapore}
  \city{Singapore}
  \country{Singapore}
}

\renewcommand{\shortauthors}{Yingzhi He et al.}

\begin{abstract}
\za{Recent advances in generative recommendation have leveraged pretrained LLMs by formulating sequential recommendation as autoregressive generation over a unified token space comprising language tokens and itemic identifiers, where each item is represented by a compact sequence of discrete tokens, namely Semantic IDs (SIDs).}
\za{This SID-based formulation enables efficient decoding over large-scale item corpora and provides a natural interface for LLM-based recommenders to leverage rich world knowledge.}
\za{Meanwhile, breakthroughs in LLM reasoning motivate reasoning-enhanced recommendation, yet effective reasoning over SIDs remains underexplored and challenging.}
\za{Itemic tokens are not natively meaningful to LLMs; moreover, recommendation-oriented SID reasoning is hard to evaluate, making high-quality supervision scarce.}

\za{To address these challenges, we propose \textbf{\ours}, a two-stage framework that elicits reasoning over SIDs by strengthening SID--language alignment to unlock transferable LLM reasoning, rather than relying on large amounts of recommendation-specific reasoning traces.}
\za{Concretely, \ours first enhances SID-language alignment via multi-task training on an enriched SID-centered corpus synthesized by a stronger teacher model, grounding itemic tokens in diverse semantic and behavioral contexts.
Building on this enhanced alignment, \ours further improves recommendation reasoning through outcome-driven reinforced optimization, which guides the model toward effective reasoning trajectories without requiring explicit reasoning annotations.}
\za{Extensive experiments on three real-world datasets demonstrate the effectiveness of our reasoning-augmented SID-based generative recommendation.
Beyond accuracy, the results highlight the broader potential of large reasoning models for generative recommendation, including improved interpretability and cross-domain generalization.}
Our codes are available at \url{https://github.com/HappyPointer/SIDReasoner}.
\end{abstract}




\maketitle
\newcommand\kddavailabilityurl{https://doi.org/10.5281/zenodo.20508816}
\ifdefempty{\kddavailabilityurl}{}{
\begingroup\small\noindent\raggedright\textbf{Resource Availability:}\\
The source code of this paper has been made publicly available at \url{\kddavailabilityurl}.
\endgroup
}

\section{Introduction}

Generative recommendation has recently emerged as a promising paradigm for sequential recommendation with an increasing attention on Semantic ID (SID)-based methods \cite{TIGER, LETTER, OneRec-v1, OneRec-v2, MiniOneRec}. 
By quantizing an item's semantic embedding into a compact sequence of discrete identifiers called Semantic IDs (SID) \cite{VQRec, TIGER}, item recommendation can be formulated as autoregressive generation task over SID sequences. 
This sequence prediction paradigm naturally aligns with the Transformer architecture, 
facilitating seamless integration of generative methods with pretrained large language models (LLMs), whose rich semantic knowledge can facilitate recommendation quality and cross-domain generalization \cite{LCRec, PLUM}.
Despite these advances, reasoning, one of the central capabilities behind recent breakthroughs of LLMs \cite{COT, wu2025inference, T1-Advancing, DeepseekR1}, remains at an early stage of exploration in generative recommendation, with its full potential yet to be uncovered.

Existing approaches to reasoning in recommendation predominantly represent items with textual descriptions, leveraging the rich world knowledge of LLMs to comprehend item semantics and infer user intentions behind interactions \cite{tsai2024leveraging, R2Rec, fang2025reason4rec, ReaRec, LatentR3, yu2025thinkrec}.
These methods can be broadly categorized into two paradigms: explicit reasoning and latent reasoning.
Explicit reasoning generates natural language rationales to explicitly articulate user preferences \cite{tsai2024leveraging, fang2025reason4rec, zeng2025optimizing, yu2025thinkrec, R2Rec, RecOne}. The latent reasoning improves efficiency by skipping explicit chain-of-thought generation, but sacrifices interpretability and controllability \cite{ReaRec, LatentR3, lin2025order, Onepiece}.
However, representing items purely with textual descriptions suffers from severe decoding inefficiency and item grounding issues, limiting its practicality in real-world systems \cite{Kuaishou_survey}.
A recent promising direction avoids such issues by representing items with SIDs and expanding SID tokens into the LLM vocabulary. After the SID-language alignment training, the model can reason in language and recommend in SIDs within a unified representation space \cite{PLUM, OneRec-think}.
While this paradigm offers new potential for reasoning in recommendation, its effectiveness critically depends on the quality of SID–language alignment, and existing successes largely rely on industrial-scale pretraining with substantial computational cost.

In this work, we investigate how to achieve effective and data-efficient reasoning for generative recommendation over SIDs.
Reasoning in recommendation is inherently challenging, as it models implicit user preferences that are not directly observable. 
This implicitness gives rise to two fundamental challenges: first, high-quality reasoning supervision is scarce and expensive to obtain. 
Second, the quality of recommendation reasoning is hard to evaluate given the implicit nature of user preferences.

\za{To tackle these challenges, we propose \textbf{\ours}, a data-efficient two-stage framework that elicits reasoning over SIDs by strengthening SID--language alignment to unlock transferable LLM reasoning.}
First, to mitigate the scarcity of high-quality recommendation reasoning supervision, we strengthen SID–language alignment as a complement alongside recommendation-specific reasoning traces, enabling general knowledge and reasoning abilities to better generalize to recommendation tasks.
Concretely, we establish SID–language grounding through multi-task alignment, where the model comprehend semantics of SID tokens through a wide range of recommendation-related data.
We further leverage a large teacher model with rich world knowledge to perform semantic expansion based on item titles, synthesizing more diverse and informative SID-centered descriptive texts.
This enriched corpus exposes the model to broader and more fine-grained semantic associations for SID tokens, strengthening their alignment with natural language.
Second, once the model attains a comprehensive understanding of SIDs and can transfer general knowledge to recommendation tasks, we enable it to self-explore reasoning patterns that are effective for recommendation.
In the absence of a direct evaluation criterion for reasoning quality, we adopt Group Relative Policy Optimization (GRPO) \cite{GRPO} to provide outcome-based feedback and steer the model toward effective reasoning trajectories.
Extensive experiments on three real-world datasets demonstrate that, with enriched SID–language alignment, effective SID alignment and reasoning can be achieved even at academic-scale dataset, enabling LLMs to perform effective reasoning-based generative recommendation in a data-efficient manner.
Beyond effectiveness, our empirical results further reveal the broader potential of reasoning in recommendation, including strong cross-domain generalization and improved interpretability through explicit reasoning processes.

We summarize our main contributions as follows:
\begin{itemize}[leftmargin=*]
    \item We investigate how to enable effective reasoning in SID-based generative recommendation, and propose a data-efficient framework that allows LLMs to reason over itemic tokens, alleviating the need for explicit reasoning supervision.
    \item We empirically demonstrate the broader potential of reasoning in recommendation beyond accuracy improvements (\cf Section \ref{sec:in_domain_perf}), including cross-domain generalization (\cf Section \ref{sec:problem_formulation}) and interpretability (\cf Section \ref{sec:case_study}).
    \item We design and release an open-sourced reasoning-based generative recommendation pipeline, facilitating future investigations into reasoning-enhanced generative recommendation.
\end{itemize}

\section{Related Work}
In this section, we briefly review the works related to this paper from two main categories: 1) generative recommendation, and 2) large reasoning models.

\begin{figure*}[t]
    \centering
    \includegraphics[width=0.98\textwidth]{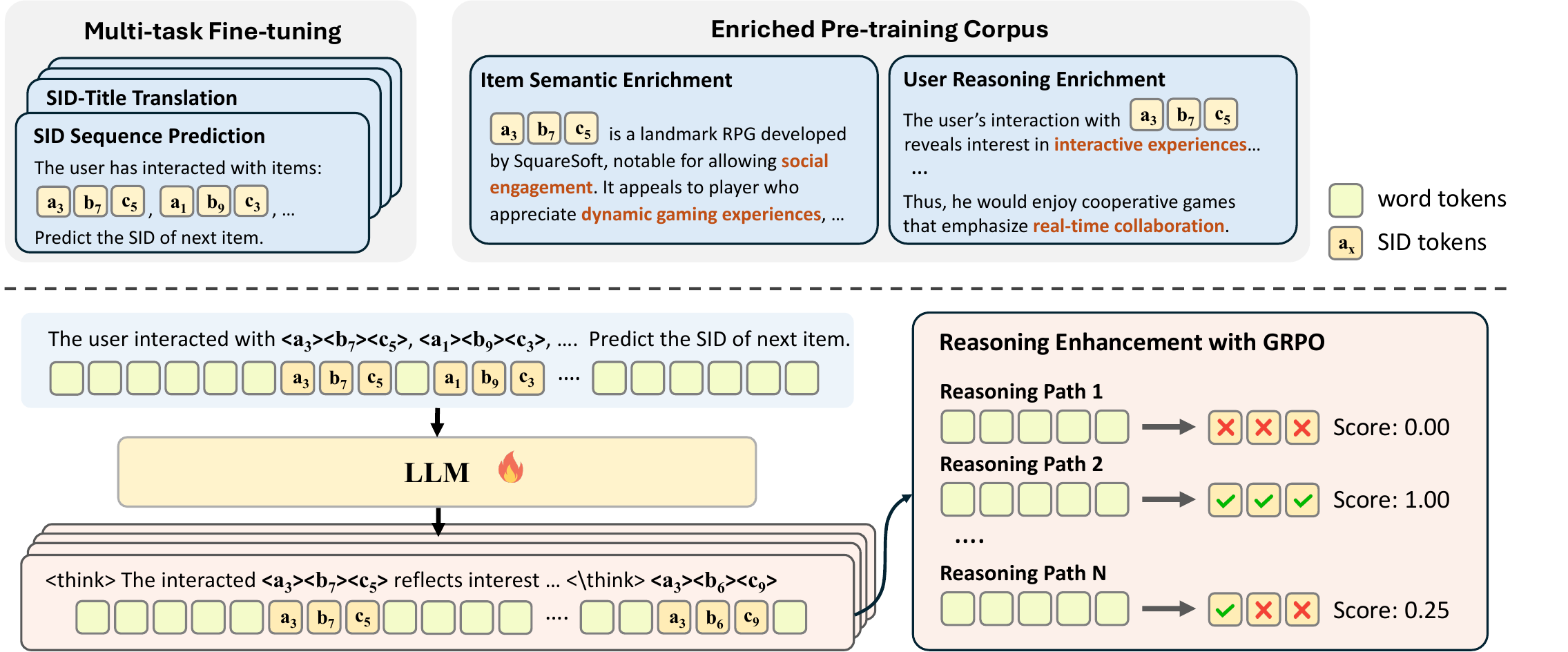}
    \vspace{-0.1in}
    \caption{Illustration of the overall framework of our proposed \ours. 
    }
    \vspace{-0.1in}
    \label{fig:framework}
\end{figure*}

\subsection{Generative Recommendation}
With the growing success of LLMs and the transformer architecture, generative recommendation has emerged as a promising research direction in the recommender systems. Existing studies in generative recommendation can be broadly classified into three categories: sparse ID-based methods, text-based methods, and semantic ID-based methods.

\noindent \textbf{Sparse ID-based methods} represent each item as a unique identifier and serialize user interactions into token sequences, formulating recommendation as next-token prediction tasks \cite{HSTU, P5, PinRec}. Existing explorationss has evolved along two major directions: incorporating additional features or contexts for controllable generation \cite{MTGR, LUM, PinRec}, and optimizing token structures to improve efficiency \cite{GenRank, DFGR}. 
However, sparse ID-based methods suffer from scalability issues due to the large output space and are ineffective in cold-start and long-tail scenarios \cite{Kuaishou_survey}.

\noindent \textbf{Text-based methods} represent each item using its natural language description and generate the textual description of the target item as the recommendation output \cite{M6-Rec, RecSysLLM, InstructRec}. 
The pretrained language models, which preserve strong semantic understanding capabilities, are further fine-tuned on recommendation tasks \cite{TallRec, BIGRec, Llara, SDPO}, exhibiting strong generalization to cold-start items and enabling more interpretable recommendations \cite{RecInterpretor, Recexplainer}.
However, generating long textual sequences also introduces substantial computational overhead and faces inherent challenges in reliably grounding the output to real-world items \cite{Llmtreerec, DecodingMatter}, making such approaches difficult to deploy in practical recommendation systems.

\noindent \textbf{Semantic ID-based methods} represent items with short sequences of discrete codes obtained by quantizing continuous item embeddings \cite{VQRec, TIGER, MiniOneRec}. Semantic IDs (SIDs) offer a trade-off between semantic expressiveness and decoding efficiency, preserving rich semantic information under a compact decoding space.
Subsequent studies further extend this paradigm by integrating collaborative signals into SID construction \cite{LETTER, EAGER, Unger, OneRec-v1, OneRec-v2} and advancing quantization schemes to improve representation quality or generation efficiency \cite{ParallelSID, CoFiRec, Pctx}.
Meanwhile, recent efforts explore the integration of Semantic IDs into pretrained language models by aligning SID tokens with the semantic space of LLMs \cite{LCRec, PLUM, OneRec-think, OpenOneRec, EAGER-LLM}. This enables models to speak in natural language as well as recommend by generating SIDs. Such integration unlocks new capabilities of generative recommendation, including interactive recommendation, explainable decision-making, and explicit reasoning over user–item interactions \cite{OneRec-think}.

\subsection{Large Reasonging Models}
In large language models, reasoning decomposes complex problems into a series of simpler steps and allows the model to answer with higher confidence through an additional reasoning process \cite{kojima2022large, zelikman2022star, snell2024scaling, 8020_rule}.
This is most commonly reflected in Chain-of-Thought (CoT) \cite{COT}, which introduces explicit token-level intermediate steps and trains or prompts the model to solve complex problems by generating these steps prior to the final answer.
The quality of reasoning can be further improved by allocating additional inference-time computation, for example through parallel sampling \cite{selfconsistency, lightman2023let, xie2023self}, and iterative refinement \cite{kamoi2024can, huang2023large, welleck2022generating}. Moreover, beyond heuristic control at inference time, reinforcement learning with verifiable rewards provides a more direct way to refine the reasoning process, improving both the reliability of intermediate inference and final correctness \cite{DeepseekR1, T1-Advancing, arora2025training, aggarwal2025l1}.
Overall, large reasoning models can be understood as systems that explicitly control and allocate additional test-time computation to achieve stronger problem-solving performance \cite{wu2025inference, AlphaReasoner, zhang2025and}.

In recommendation systems, reasoning is similarly introduced to replace one-shot scoring with multi-step inference processes that progressively refine user intent and recommendation decisions. Existing explorations of incorporating reasoning into recommendation tasks can be broadly categorized into two directions: explicit reasoning and latent reasoning. 
The first direction, explicit reasoning, relies on explicit natural language rationales as intermediate inference steps, enabling interpretable and controllable multi-step recommendation \cite{tsai2024leveraging, fang2025reason4rec, zeng2025optimizing, yu2025thinkrec, R2Rec, RecOne, OneRec-think}.
In contrast, the second direction, latent reasoning, performs multi-step inference by iteratively refining the model’s latent states in the continuous representation space, without relying on explicit natural language rationales \cite{LatentR3, ReaRec, lin2025order, Onepiece}.
Despite recent progress, both directions still face substantial challenges. Explicit reasoning relies on high-quality recommendation-oriented reasoning data that is not naturally available, whereas latent reasoning lacks interpretability and controllability in the decision process.

\section{Methodoly}
In this section, we present our reasoning-enhanced framework for generative recommendation, as illustrated in Figure~\ref{fig:framework}. We first formalize the problem, then describe how we quantize items and align itemic tokens with natural language through multi-task training and enriched semantic corpus. Finally, we present reinforced optimization that enhances the model’s reasoning ability on recommendation.

\subsection{Task Formulation} \label{sec:problem_formulation}
\subsubsection{Tokenization with Semantic IDs.}
\za{
Generative recommendation aims to generate the next item that a user is likely to interact with.
Let $\mathcal{U}$ and $\mathcal{I}$ denote the sets of users and items.
For each user $u \in \mathcal{U}$, a chronological interaction history is observed as
$\mathcal{H}_u = (i_1, i_2, \ldots, i_T)$, where $i_t \in \mathcal{I}$, and the goal is to generate the next item $i_{T+1}$ conditioned on $\mathcal{H}_u$.
To leverage a pretrained autoregressive LLM for recommendation, each item $i$ is represented as a sequence of discrete tokens that can be processed by next-token prediction.
Besides the LLM language-token vocabulary $\mathcal{V}_{\text{LM}}$, a recommendation-specific itemic token vocabulary $\mathcal{S}$ is defined, and each item $i \in \mathcal{I}$ is mapped to a fixed-length Semantic ID (SID) sequence:
$\mathrm{SID}(i) = (s_i^1, s_i^2, \ldots, s_i^L), s_i^l \in \mathcal{S},$
where $L$ is the SID length.
The SID sequence is obtained by encoding the item metadata $t_i$ (e.g., title, category, and optionally a brief description) into a continuous semantic embedding and then quantizing it into $L$ discrete semantic tokens, yielding a compact tokenization that preserves item semantics.
}

\subsubsection{Generative Recommendation with Reasoning.}
\za{Under this formulation, generating the next item $i_{T+1}$ corresponds to autoregressively generating its $L$ SID tokens from $\mathcal{S}$, \ie one item generation is realized by $L$ consecutive next-token prediction steps over the itemic vocabulary.}
\za{The token space of the LLM is augmented to include both natural language tokens and itemic tokens, i.e., $\mathcal{V} = \mathcal{V}_{\text{LM}} \cup \mathcal{S}$, enabling textual reasoning and item generation within a unified autoregressive model.
For each user $u$, the interaction history $\mathcal{H}_u=(i_1,i_2,\ldots,i_T)$ is converted into itemic token sequences by mapping each interacted item $i_t$ to its SID sequence
$\mathbf{y}_t=\mathrm{SID}(i_t)\in\mathcal{S}^L$.}
\za{The history is then represented as a flattened itemic-token context
$\mathcal{H}_u=\mathrm{concat}(\mathbf{y}_1,\mathbf{y}_2,\ldots,\mathbf{y}_T)$.
Together with an instruction-style prompt $\mathbf{p}$, the model input is formed as
$\mathcal{C}_u = [\mathbf{p}; \mathcal{H}_u]$.}
Conditioned on this context \za{$\mathcal{C}_u$}, the \za{recommender} first generates an intermediate reasoning sequence \(\tau = (r_1, r_2, \ldots, r_M)\), and subsequently generates the next item \(\mathbf{y}_{T+1} = \mathrm{SID}(T+1)\).
The overall generation process is modeled autoregressively as
\begin{equation}
\tau \sim \pi_\theta(\cdot \mid \mathcal{C}_u), \quad
\mathbf{y}_{T+1} \sim \pi_\theta(\cdot \mid \mathcal{C}_u, \tau),
\end{equation}
where \(\pi_\theta\) denotes the LLM-based generative policy. In this formulation, the reasoning sequence $\tau$ serves as an explicit intermediate inference step that guides the generation of the next item based on the user’s historical behavior.

\subsection{Enriched SID-Language Alignment}

\subsubsection{Item Quantization}

To obtain discrete Semantic IDs from item textual metadata, we adopt an item quantization based RQ-VAE \cite{RQ-VAE}, which has become a widely adopted tokenization method in generative recommendation \cite{TIGER}.
Specifically, for each item $i \in \mathcal{I}$, we first encode its textual metadata $t_i$ into a continuous semantic representation $\mathbf{z}_i \in \mathbb{R}^d$ using a text encoder. The embedding is then fed into a quantization module, where a sequence of discrete codes is produced through a multi-stage residual quantization process. Concretely, we maintain $L$ codebooks $\{\mathcal{C}^1, \mathcal{C}^2, \ldots, \mathcal{C}^L\}$, where each codebook $\mathcal{C}^l = \{\mathbf{e}^l_1, \ldots, \mathbf{e}^l_K\}$ contains $K$ code vectors.
At quantization stage $l$, the residual vector $\mathbf{r}^{l-1}$ is approximated by selecting the nearest codeword from $\mathcal{C}^l$, and the residual is updated accordingly:
\begin{equation}
s_i^l = \arg\min_{k} \|\mathbf{r}^{l-1} - \mathbf{e}^l_k\|_2^2, 
\quad
\mathbf{r}^{l} = \mathbf{r}^{l-1} - \mathbf{e}^l_{s_i^l},
\end{equation}
with $\mathbf{r}^0 = \mathbf{z}_i$. After $L$ stages, the item embedding is represented as the sum of the selected codewords across all levels.
The resulting discrete indices $(s_i^1, s_i^2, \ldots, s_i^L)$ constitute the Semantic ID sequence $\mathrm{SID}(i)$ of item $i$.
The codebooks and the quantization module are optimized with a joint loss $\mathcal{L}_{\text{RQ-VAE}}$ consisting of a reconstruction term $\mathcal{L}_{\text{recon}}$ and a residual quantization regularization term $\mathcal{L}_{\text{RQ}}$. Specifically, 
\begin{equation}
\mathcal{L}_{\text{RQ-VAE}} = \mathcal{L}_{\text{recon}} + \mathcal{L}_{\text{RQ}},
\end{equation}
\begin{equation}
\mathcal{L}_{\text{recon}} =
\|\mathbf{z}_i - \hat{\mathbf{z}}_i\|_2^2,
\end{equation}
\begin{equation}
\mathcal{L}_{\text{RQ}} =
\sum_{l=1}^{L}
\left(
\|\text{sg}[\mathbf{r}^{l-1}] - \mathbf{e}^l_{s_i^l}\|_2^2
+
\beta \|\mathbf{r}^{l-1} - \text{sg}[\mathbf{e}^l_{s_i^l}]\|_2^2
\right).
\end{equation}
where $\hat{\mathbf{z}}_i$ denotes the reconstructed embedding obtained from the decoder, $\text{sg}[\cdot]$ denotes the stop-gradient operator and $\beta$ controls the strength of the commitment term.
The residual quantization scheme allows each item to be represented by a compact sequence of discrete SID tokens while preserving semantic fidelity, and enables efficient modeling within an autoregressive language model.

\subsubsection{Multi-task Fine-tuning}
While SIDs provide a compact discrete representation of items, the corresponding SID tokens are newly introduced into the LLM vocabulary with randomly initialized embeddings, and therefore are semantically meaningless to the model before fine-tuning.
To enable the model to understand, reason over, and generate SID tokens meaningfully in the recommendation context, we conduct a multi-task fine-tuning stage where SID tokens and natural language tokens co-occur across diverse recommendation scenarios. Through these tasks, the model jointly learns recommendation signals from SID sequences and the semantic correspondence between SID tokens and textual item descriptions, thereby achieving effective SID–language alignment.
To this end, we adopt multiple recommendation-oriented tasks under a unified autoregressive objective, which can be grouped into two categories.
\begin{itemize}[leftmargin=*, itemsep=2pt]
    \item \textbf{Item prediction.}
    Given a user’s historical interactions, the model is trained to predict future interactions, where items are represented with either SID sequences or textual descriptions. This task enables the model to capture behavioral patterns while grounding SID tokens in semantic item information.
    \item \textbf{SID translation.}
    To facilitate SID-language alignment, the model is further fine-tuned to translate SIDs, where the model generates the textual title from a given SID sequence and, conversely, the SID sequence from its textual title. This task encourages the model to associate SID tokens with their underlying semantic meanings in the language space.
\end{itemize}
Due to space limitations, detailed descriptions of each task are provided in appendix~\ref{appendix:alignment_formats}. All tasks are jointly optimized under the standard next-token prediction objective, enabling effective SID–language alignment for generative recommendation.

\begin{figure}[t]
    \centering
    \includegraphics[width = 0.8\linewidth]{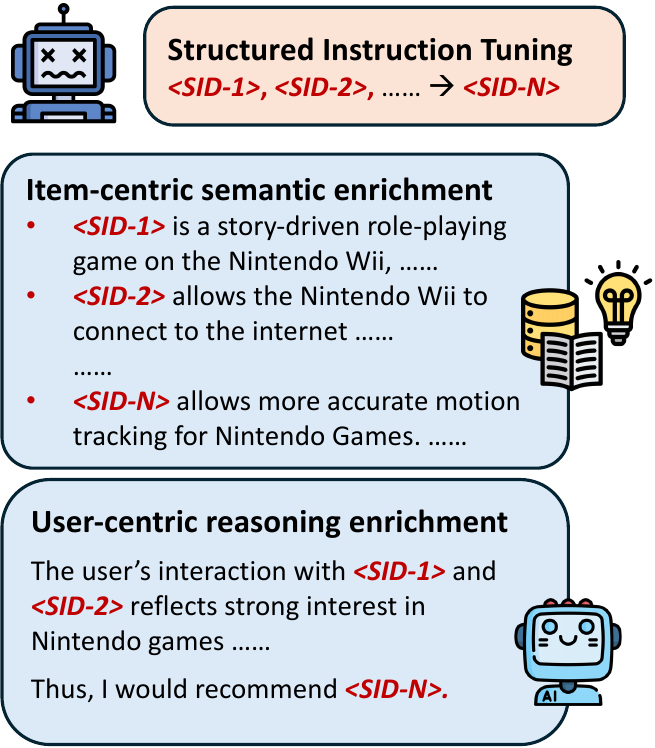}
    \vspace{-0.12in}
    \caption{Illustration of enriched alignment corpus.}
    \label{fig:method_data_enrich}
    \vspace{-0.2in}
\end{figure}

\subsubsection{Enriched Corpus Pre-training}
Although the recommender benefits from multi-task training, the limited diversity of task formats leads to restricted and repetitive alignment between SID tokens and word tokens, which results in insufficient understanding of SID semantics and reduces the quality of item-related reasoning.
To address this issue, we leverage a large teacher language model with rich world knowledge to synthesize a more semantically enriched item–language corpus for additional pre-training, as illustrated in Figure~\ref{fig:method_data_enrich}. The goal is to provide more diverse and informative SID–language associations, thereby strengthening the model’s understanding of SID semantics. In particular, we augment the existing recommendation data from two perspectives.
\begin{itemize}[leftmargin=*, itemsep=2pt]
    \item \textbf{Item-centric semantic enrichment.}
    For each item, the teacher model first expands its textual metadata into a structured semantic description, including attributes like usage scenarios, target users, key characteristics, \etc and then generates a coherent paragraph that interleaves SID tokens with natural language. This produces training samples where SID tokens appear in semantically rich and linguistically diverse contexts, strengthening their alignment with underlying meanings.
    \item \textbf{User-centric reasoning enrichment.}
    Similarly, for each user interaction sequence, the teacher model infers the underlying motivations and generates a concise mixed SID–language description that summarizes user's historical behavior, reasoning over motivations, and the resulting actions.
    These enriched samples strengthen SID–language alignment and, more importantly, associate SID tokens with user behavior reasoning.
\end{itemize}

In addition, to prevent the model from overfitting to recommendation tasks and degrading its general language and reasoning abilities, we further mix the general-domain reasoning data into the alignment corpus. This auxiliary general data helps preserve the model’s original general reasoning ability while learning enriched SID–language associations.

\subsection{Reinforced Reasoning Enhancement}

\subsubsection{Cold-start Reasoning Activation}
Although the model acquires the ability to jointly reason and recommend through enriched SID–language alignment, it does not necessarily default to generating explicit reasoning before itemic predictions at inference time.
To reliably activate the reasoning behavior, we introduce a lightweight cold-start reasoning activation stage.
Specifically, we use teacher-generated reasoning from the alignment stage to construct supervised samples, and apply standard supervised fine-tuning to enforce a reason-then-recommend generation pattern, where the model first produces a reasoning sequence $\tau$ and then predicts the target SID sequence $\mathbf{y}_{T+1}$ conditioned on the user context $\mathcal{C}_u$.

It is worth noting that, due to prior SID–language alignment, the model already possesses the capability to perform reasoning-based recommendation.
This stage mainly serves to improve the reliability of response formatting, ensuring that reasoning is consistently generated before recommendation.
In practice, this activation step is lightweight and takes only a single epoch of fine-tuning.

\subsubsection{Group-wise Reinforcement Learning}

After reasoning activation, the model has a more stable initialization for reasoning in recommendation. We further refine the model’s policy by reinforcement learning on recommendation tasks, where the policy $\pi_\theta$ is directly optimized with rewards from recommendation accuracy.

To this end, we define the reward for a reasoning sequence $\tau$ and predicted item representation $\mathbf{y}$ as
\begin{equation}
R_\theta(\tau, \mathbf{y}) = R_{sr}(\mathbf{y}, i_{T+1}) + \lambda \, R_f(\mathbf{y}),
\end{equation}
where $\lambda$ is a balancing coefficient, $R_{sr}(\cdot)$ and $R_f(\cdot)$ are the stepwise rule-based reward and format reward, respectively.
The stepwise rule-based reward $R_{sr}(\mathbf{y}, i_{T+1})$ assesses the prediction quality by comparing the generated item representation $\mathbf{y}$ with the ground-truth next item. Let $L$ be the length of the item representation and $m$ be the length of the longest correct prefix between $\mathbf{y}$ and the ground-truth $\mathbf{y}_{T+1}$, then
\begin{equation}
R_{sr}(\mathbf{y}, i_{T+1}) = \frac{1}{2^{\,L-m}},
\end{equation}
which provides a smoothly increasing reward as more tokens in $\mathbf{y}$ match the ground truth, approaching 1 when the entire sequence is correct.
The format reward $R_f(\mathbf{y})$ checks if the prediction $\mathbf{y}$ is structurally valid and corresponds to an existing item:
\begin{equation}
R_f(\mathbf{y}) =
\begin{cases}
1, & \text{if } \mathbf{y} \text{ maps to a catalog-existing item},\\
0, & \text{otherwise},
\end{cases}
\end{equation}
which encourages the model to generate valid and meaningful itemic token sequences.

With the reward function defined, we optimize the policy $\pi_\theta$ using Group Relative Policy Optimization (GRPO)~\cite{GRPO}.
For each user context $\mathcal{C}_u$, we sample a group of $K$ reasoning–prediction trajectories
$\{o^k\}_{k=1}^K$ from the previous policy $\pi_{\theta_{\text{old}}}$,
where each trajectory $o^k = \tau^k \circ \mathbf{y}^k$ consists of a reasoning sequence followed by a SID prediction.
Each trajectory is assigned an outcome-based reward $R^k$.
GRPO performs group-wise normalization of rewards and updates the policy by maximizing the following clipped surrogate objective:
\begin{equation}
\small
\begin{aligned}
\mathcal{L}_{\text{GRPO}}(\theta)
=&\;
\mathbb{E}_{\mathcal{C}_u}
\Bigg[
\frac{1}{K}\sum_{k=1}^{K}
\min\!\big(
\rho^k(\theta)\hat{A}^k,\,
\text{clip}(\rho^k(\theta),1-\eta,1+\eta)\hat{A}^k
\big)
\Bigg] \\
&\;-\beta\, D_{\mathrm{KL}}\!\big(\pi_\theta \,\|\, \pi_{\text{ref}}\big),
\end{aligned}
\end{equation}
where $\rho^k(\theta)=\frac{\pi_\theta(o^k\mid\mathcal{C}_u)}{\pi_{\theta_{\text{old}}}(o^k\mid\mathcal{C}_u)}$
is the trajectory-level importance ratio,
$\hat{A}^k$ denotes the group-normalized advantage from $\{R^j\}_{j=1}^K$,
$\eta$ is the clipping threshold, and $\beta$ controls KL regularization.

By optimizing this objective, the model is encouraged to increase the likelihood of reasoning–prediction trajectories $o^k$ that yield higher relative reward, while suppressing less effective reasoning paths.
This reinforced optimization progressively refines both the quality of generated reasoning and the accuracy of item prediction.

\begin{table}[t]
\begin{center}
\caption{The statistics of recommendation datasets.}
\label{tab:dataset_satistics}
\vspace{-0.1in}
\resizebox{0.48\textwidth}{!}{
    \begin{tabular}{lcccccc}
        \toprule
        \textbf{Dataset} & \textbf{\#Items}  & \textbf{\#Interactions} & \textbf{Train} & \textbf{Val\&Test}\\
        \midrule
         Games & 3,858 & 61,417 & 49,133 & 6,142 \\
         Office & 3,459 & 48,656 & 38,924 & 4,866 \\
         Industrial & 3,686 & 45,325 & 36,259 & 4,533 \\
        \bottomrule
    \end{tabular}
}
\end{center}
\vspace{-0.15in}
\end{table}

\section{Experiments}
In this section, we present the experimental results and corresponding analysis to answer the following research questions (\textbf{RQ}s).
\begin{itemize}[leftmargin=*]
    \item \textbf{RQ1:} How effective is our proposed \ours?
    \item \textbf{RQ2:} How does each part contribute to model performance?
    \item \textbf{RQ3:} What are the key properties of \ours?
\end{itemize}

\subsection{Experiment Settings}

\subsubsection{Datasets and Evaluation.}

Following prior works \cite{DecodingMatter, BIGRec, SDPO}, we conduct experiments on three real-world datasets collected from Amazon platform \cite{Amazon23}: Video Games (Games), Office Products (Office), and Industrial and Scientific (Industrial). The datasets contain user interaction histories with each interacted item is associated with rich textual metadata.
For all datasets, we first apply 5-core filtering to ensure that each user and each item has at least five interactions. To control sequence length and maintain a consistent modeling setting, we truncate each user’s historical interaction sequence using the sliding window with maximum length set to 10.
We adopt a temporal split to simulate realistic recommendation scenarios. For each user, interactions are sorted chronologically and divided into training, validation, and test sets with a ratio of 8:1:1, where the most recent interactions are reserved for validation and testing. This time-aware partition ensures that the model is always evaluated on future behaviors. Detailed dataset statistics of each category are presented in Table~\ref{tab:dataset_satistics}. The datasets span three different domains on Amazon platform with comparable scales in terms of the number of items and interactions.

For evaluation metrics, we adopt Recall@K and NDCG@K with cutoff values $K \in \{5, 10\}$. We follow the full-item ranking setting, where ranking metrics are computed over the entire item set rather than sampled negatives, which is more closely aligned with real-world recommendation scenarios.

\subsubsection{Baselines.}

To validate the effectiveness of our method, we compare \ours against three different categories of baselines:
(1) traditional discriminative sequential recommenders, including
\textit{Caser} \cite{Caser}, \textit{GRU4Rec} \cite{GRU4Rec}, and \textit{SASRec} \cite{SASRec}, which model user behaviors from interaction sequences in a discriminative manner.
(2) generative recommendation methods, including \textit{TIGER} \cite{TIGER}, \textit{HSTU} \cite{HSTU}, \textit{LETTER} \cite{LETTER}, and \textit{LCRec} \cite{LCRec}
(3) Reasoning-based recommendation methods, including \textit{ReaRec} \cite{ReaRec} which leverage latent reasoning, and $R^{2}ec$ \cite{R2Rec}, which incorporate explicit textual reasoning into the recommendation process.
Detailed descriptions of each baseline method are presented in Appendix~\ref{appendix:baselines}.

\subsubsection{Implementation Details.}
We adopt Qwen3-1.7B \cite{Qwen3} as the backbone model and perform full-parameter fine-tuning throughout all stages. SID tokens are appended to the tokenizer vocabulary with randomly initialized embeddings. During SID–language alignment training, we use the AdamW optimizer \cite{AdamW} with a batch size of 1024 and apply early stopping based on the model’s performance on predicting the groundtruth SID given historical SID sequences. For enriched corpus construction, we call the GPT4o-mini API to synthesize semantically enriched SID–language data. More detailed descriptions of the training data are provided in Appendix~\ref{appendix:alignment_formats} and Appendix~\ref{appendix:enrich_prompts}.

During enriched SID-language alignment, we apply early stopping with patience to 2 and select the checkpoint with the lowest evaluation loss for the subsequent reasoning activation training. The reasoning activation stage is trained for only one epoch. This stage is not intended to teach the model recommendation reasoning from scratch, since the corresponding ability has already been established during enriched SID-language alignment. The reasoning activation mainly aims to improve instruction following and stabilize the response format, ensuring that the model consistently generates valid reasoning before making SID-based recommendations. This provides a reliable initialization for the subsequent reinforcement learning.

In the reinforcement learning stage, we implement GRPO based on verl \cite{verl}. We set the rollout number to 16, the KL regularization coefficient to \(1\times10^{-3}\), and the batch size to 256. The weighting coefficient $\lambda$ before the format reward is set to 0.1, and the learning rate is fixed at \(5\times10^{-7}\).

\subsection{Performance Comparison (RQ1)}

\subsubsection{Reasoning Enhances In-Domain Performance.}

\begin{table*}[t]
\caption{Performance comparison over different baseline methods. R is shorts for Recall, N is short for NDCG.}
\label{tab:overall_performance}
\centering
\setlength{\tabcolsep}{1mm}{
    \resizebox{0.93\textwidth}{!} {
        \begin{tabular}{ll cccc cccc cccc}
        \toprule

        \multirow{2.4}{*}{Models} & \multicolumn{4}{c}{\textbf{Games}} & \multicolumn{4}{c}{\textbf{Office}} & \multicolumn{4}{c}{\textbf{Industrial}} \\
        \cmidrule(lr){2-5} \cmidrule(lr){6-9} \cmidrule(lr){10-13} & R@5 & N@5 & R@10 & N@10 & R@5 & N@5 & R@10 & N@10 & R@5 & N@5 & R@10 & N@10 \\
        \midrule

        Caser & 0.0376 & 0.0241 & 0.0659 & 0.0332 & 0.0880 & 0.0663 & 0.1114 & 0.0738 & 0.0664 & 0.0528 & 0.0852 & 0.0588 \\
        GRU4Rec & 0.0329 & 0.0219 & 0.0599 & 0.0305 & 0.0682 & 0.0480 & 0.0974 & 0.0574 & 0.0788 & 0.0578 & 0.1030 & 0.0649 \\
        SASRec & 0.0501 & 0.0345 & 0.0723 & 0.0416 & 0.1019 & 0.0824 & 0.1167 & 0.0871 & 0.0807 & 0.0647 & 0.0964 & 0.0697 \\
        \midrule
        TIGER & 0.0489 & 0.0300 & 0.0763 & 0.0402 & 0.1270 & 0.1037 & 0.1429 & 0.1121 & 0.1003 & 0.0823 & 0.1325 & 0.0924 \\
        HSTU & 0.0539 & 0.0396 & 0.0746 & 0.0462 & 0.1204 & 0.1069 & 0.1323 & 0.1107 & 0.1008 & \underline{0.0898} & 0.1138 & 0.0940 \\
        LETTER & 0.0445 & 0.0294 & 0.0709 & 0.0378 & \underline{0.1315} & \underline{0.1074} & \underline{0.1520} & \underline{0.1139} & \underline{0.1080} & 0.0850 & \underline{0.1389} & \underline{0.0950} \\
        LCRec & 0.0441 & 0.0274 & 0.0876 & 0.0412 & 0.0964 & 0.0699 & 0.1487 & 0.0867 & 0.0805 & 0.0520 & 0.1330 & 0.0687 \\

        \midrule
        ReaRec & 0.0568 & 0.0381 & 0.0843 & 0.0470 & 0.1173 & 0.0988 & 0.1385 & 0.1057 & 0.0973 & 0.0796 & 0.1205 & 0.0870 \\
        R\textsuperscript{2}ec & \underline{0.0655} & \underline{0.0399} & \underline{0.0931} & \underline{0.0525} & 0.1147 & 0.0894 & 0.1486 & 0.1004 & 0.0880 & 0.0774 & 0.1253 & 0.0774 \\
        
        \midrule
        \textbf{\ours} & \textbf{0.0710} & \textbf{0.0460} & \textbf{0.1031} & \textbf{0.0563} & \textbf{0.1373} & \textbf{0.1119} & \textbf{0.1648} & \textbf{0.1208} & \textbf{0.1109} & \textbf{0.0905} & \textbf{0.1438} & \textbf{0.1010} \\
        \bottomrule
        \end{tabular}
    }
}
\end{table*}

\label{sec:in_domain_perf}
We present the performance of our proposed \ours and compare it with other baseline methods on three datasets, with the results recorded in Table~\ref{tab:overall_performance}. As illustrated by the experiment results, generative recommendation methods generally outperform traditional sequential recommendation approaches across all datasets. This trend confirms the effectiveness of modeling recommendation as sequence generation over SIDs.
Moreover, after reinforcement learning, our method achieves the strongest SID prediction performance, surpassing not only conventional generative recommenders but also reasoning-enhanced recommendation approaches. These results indicate that effective reasoning over SIDs can be successfully learned and exploited to improve recommendation quality.

An interesting observation is that the effectiveness of reasoning varies notably across different item categories, leading to uneven performance gains on different datasets. Specifically, on the Games dataset, where items are rich in semantic information and align well with the world knowledge encoded in LLMs, reasoning-based methods yield substantial improvements over non-reasoning counterparts.
In contrast, on the Industrial dataset, where LLMs possess relatively limited domain-relevant knowledge, the benefits brought by reasoning are notably more limited.
The effectiveness of reasoning varies across datasets, and this phenomenon is not unique to our method. A highly consistent trend is also observed in R\textsuperscript{2}ec, which performs explicit reasoning before recommending. This trend suggests that the effectiveness of reasoning is closely tied to the availability and quality of semantic knowledge that can be leveraged by LLMs.

\subsubsection{Reasoning Generalizes to Out-of-Domain Items.}
\begin{figure}[t]
    \centering
    \includegraphics[width = 0.98\linewidth]{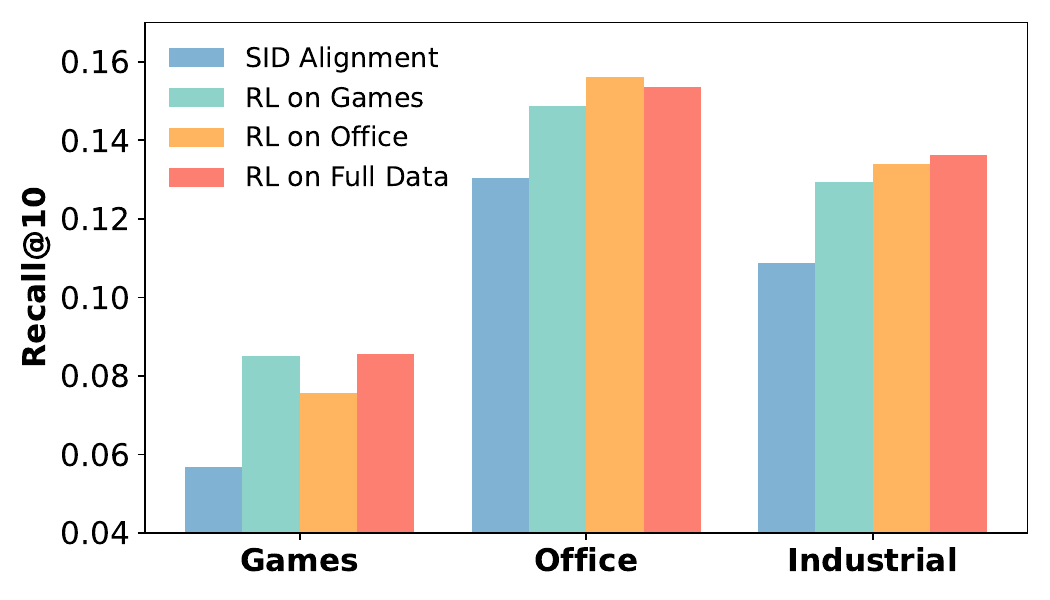}
    \caption{Performance comparison on cross-domain recommendation setting.}
    \label{fig:perf_crossdomain}
    \vspace{-0.23in}
\end{figure}

Figure~\ref{fig:perf_crossdomain} illustrates the cross-domain generalization ability of reasoning-based recommendation.
Specifically, we construct a unified SID space covering Games, Office, and Industrial datasets, and perform SID-language alignment on the mixed corpus spanning all three domains.
After the model acquires a shared semantic understanding of SIDs across domains, we conduct reasoning-oriented reinforcement learning on single datasets (Games and Office) as well as full datasets.
As shown in Figure~\ref{fig:perf_crossdomain}, RL conducted on a single domain consistently improves reasoning effectiveness on both in-domain and out-of-domain datasets.
This observation suggests that the learned reasoning capability is not tied to domain-specific item distributions, but instead captures a more general knowledge of how to reason effectively in recommendation tasks.
Such reasoning knowledge, once learned, can be transferred across domains.

\label{sec:out_of_domain_perf}

\subsection{Ablation Study (RQ2)}

\subsubsection{Alignment Strategies Affect Reasoning Effectiveness}
In this subsection, we investigate how different SID-language alignment strategies affect the reasoning potential of recommendation models, with the goal of enabling more effective reasoning on recommendation and guiding the model toward a better convergence point after reinforcement learning.
To this end, we conduct a systematic comparison of four backbone settings with increasingly diverse training corpora.
(1) \textbf{Vanilla Qwen3-1.7B}, where we directly perform reasoning activation via supervised fine-tuning on a pretrained language model, forcing it to learn recommendation reasoning over SID tokens without any explicit alignment stage.
(2) \textbf{Multi-task Alignment}, where Qwen3-1.7B is first trained with a multi-task SID--language alignment objective to establish basic semantic grounding of SID tokens, followed by reasoning activation.
(3) \textbf{Enriched Alignment}, which further augments the multi-task alignment stage with our generated enriched recommendation corpus, aiming to expose the model to more diverse and semantically rich SID-language interactions before reasoning activation.
(4) \textbf{Enriched + General Reasoning}, where we additionally incorporate general reasoning data into the alignment stage, alleviating catastrophic forgetting of the model’s general language abilities induced by recommendation-specific optimization.
For these four backbone models, we evaluate their behavior from three perspectives: the upper bound of recommendation reasoning capability, the converged performance after reinforcement learning, and general abilities.

\begin{figure}[t]
    \centering
    \includegraphics[width = 0.98\linewidth]{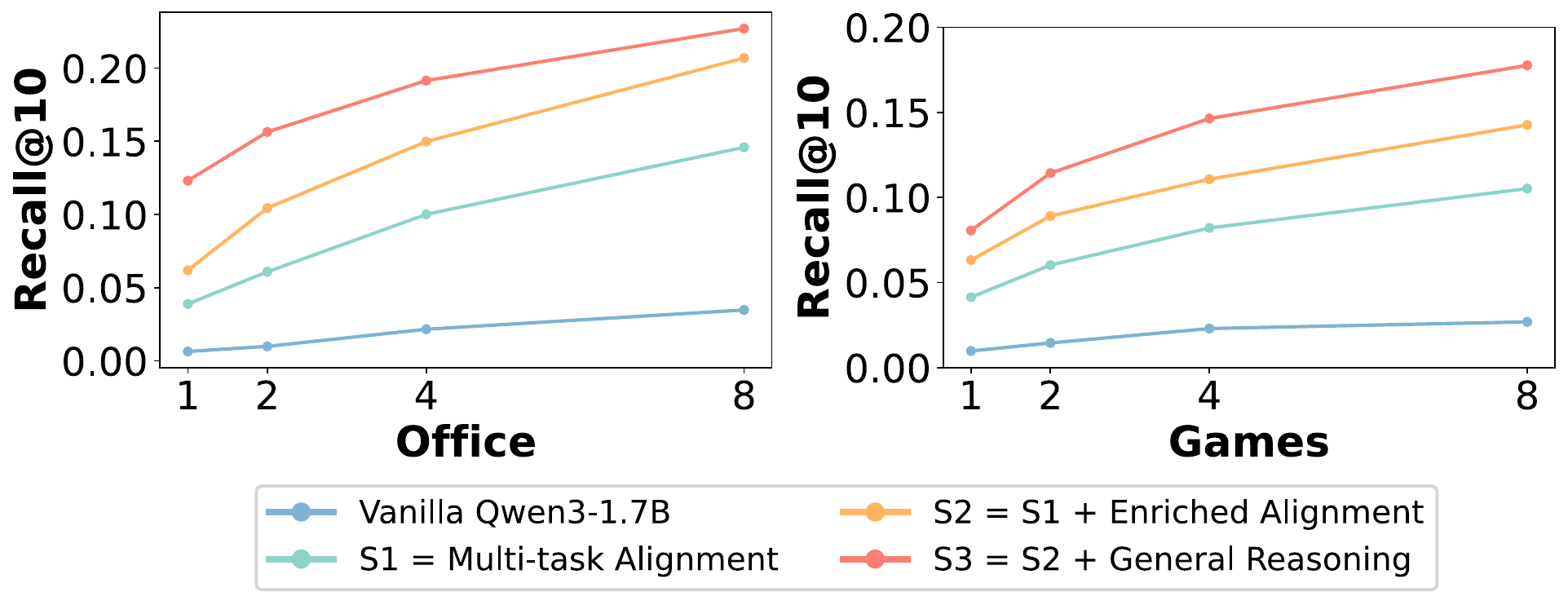}
    \vspace{-0.12in}
    \caption{Ablation performance of different alignment strategies with Best-of-$N$ reasoning selection.}
    \label{fig:ablation_bon}
\end{figure}

Figure~\ref{fig:ablation_bon} presents the Best-of-$N$ reasoning performance of the four backbone models under different alignment strategies.
The horizontal axis denotes the number of sampled reasoning trajectories $N$, while the vertical axis reports the corresponding recommendation performance measured by Recall@10.
To compute best-of-$N$ performance, we randomly sample $N$ reasoning outputs for each input, select the best reasoning trajectory according to the ground-truth item, and then evaluate the recommendation quality induced by this selected reasoning.
The best-of-$N$ performance reflects the upper bound of model's reasoning capability. 

We reports the converged performance of different backbones after RL training in Table~\ref{tab:ablation_study_rl_perf}. The experiment results suggest that models with higher best-of-$N$ performances consistently converge to better final performance. This alignment between best-of-$N$ performance and post-RL convergence results suggests that Best-of-$N$ evaluation serves as an effective indicator of a model’s reasoning capacity and its optimization potential under reinforcement learning. 
The results in both Figure~\ref{fig:ablation_bon} and Table~\ref{tab:ablation_study_rl_perf} indicate that explicit SID--language alignment is a necessary prerequisite for effective reasoning-based recommendation. Pretrained language models lack inherent knowledge of recommendation semantics and item identifiers, making direct reasoning activation insufficient for understanding SID and the recommendation task.
Multi-task alignment provides essential semantic grounding, which substantially expands the model’s reasoning capacity.
Further enriching the SID-language alignment corpus helps LLMs understands the SIDs better, and diversifies the associations between language and itemic tokens. 
Finally, mixing general reasoning data during alignment alleviates catastrophic forgetting of general language abilities and simultaneously enhances reasoning effectiveness on recommendation tasks.

\begin{table}[t]
\begin{center}
\caption{Ablation performance of different alignment strategies before and after reinforcement learning optimization.}
\label{tab:ablation_study_rl_perf}
\vspace{-0.1in}
\resizebox{0.49\textwidth}{!}{
    \begin{tabular}{lcccccc}
        \toprule
         \multirow{2.4}{*}{Models} & \multicolumn{2}{c}{Games} & \multicolumn{2}{c}{Office} \\
         \cmidrule(lr){2-3} \cmidrule(lr){4-5} & R@10 & N@10 & R@10 & N@10 \\
        \midrule
        S1 = Multi-task Alignment & 0.0414 & 0.0206 & 0.0388 & 0.0201 \\
        - After RL & 0.0741 & 0.0355 & 0.1539 & 0.0970 \\
        \midrule
        S2 = S1 + Enriched Alignment & 0.0632 & 0.0320 & 0.0619 & 0.0363 \\
        - After RL & 0.0957 & 0.0470 & 0.1607 & 0.1178 \\
        \midrule
        S3 = S2 + General Reasoning & 0.0806 & 0.0450 & 0.1231 & 0.0883 \\
        - After RL & 0.1031 & 0.0563 & 0.1648 & 0.1208\\

        \bottomrule
    \end{tabular}
}
\end{center}
\end{table}

\begin{table}[t]
\begin{center}
        \caption{Performance comparison over general abilities.}
\label{tab:ablation_general}
\vspace{-0.15in}
\resizebox{0.49\textwidth}{!}{
    \begin{tabular}{lcccccc}
        \toprule
         \textbf{Models} & \textbf{MMLU} & \textbf{IFEVAL} & \textbf{GSM8K} \\
        \midrule
        Vanilla Qwen3-1.7B & 0.6085 & 0.1793 & 0.6850 \\
        S1 = Multi-task Alignment & 0.2760 & 0.0906 & 0.0060 \\
        S2 = S1 + Enriched Alignment & 0.4464 & 0.0739 & 0.0330  \\
        S3 = S2 + General Reasoning & 0.5580 & 0.1497 & 0.5430 \\
        \bottomrule
    \end{tabular}
}
\end{center}
\vspace{-0.15in}
\end{table}

\subsubsection{Alignment Strategies Affect General Abilities}
Table~\ref{tab:ablation_general} reports the general language performance of different backbone models under various alignment strategies and compares them with the vanilla Qwen3-1.7B model.
The evaluated benchmarks include IFEval \cite{IFEVAL}, MMLU \cite{MMLU1, MMLU2}, and GSM8K \cite{GSM8K}, which assess instruction following, general reasoning, and mathematical reasoning abilities, respectively.
Overall, training on recommendation tasks leads to inevitable degradation of general language abilities, as recommendation tasks are naturally out-of-domain samples for general language models. 
This drop in general performance is most pronounced on mathematical benchmarks, where performance degrades substantially when only multi-task alignment is applied.
Enriching the alignment corpus mitigates this degradation but does not prevent it.
Incorporating general reasoning data during alignment effectively alleviates catastrophic forgetting, preserving general language abilities that are also important to effective reasoning on recommendation tasks.

\subsection{Model Study (RQ3)}

\begin{table}[t]
\begin{center}
\caption{Performance comparison over different teacher models for corpus enrichment on Games dataset.}
\label{tab:ablation_teacher}
\vspace{-0.1in}
\resizebox{0.26\textwidth}{!}{
    \begin{tabular}{lcc}
        \toprule
        \textbf{Enrichment} & \textbf{R@10} & \textbf{N@10} \\
        \midrule
        Metadata & 0.0703 & 0.0466 \\
        Qwen3-8B & 0.0727 & 0.0480 \\
        Qwen3-32B & 0.0858 & 0.0498 \\
        GPT-4o-mini & 0.1031 & 0.0563 \\
        \bottomrule
    \end{tabular}
}
\end{center}
\end{table}

\begin{figure*}[t]
    \centering
    \includegraphics[width = 1.0\linewidth]{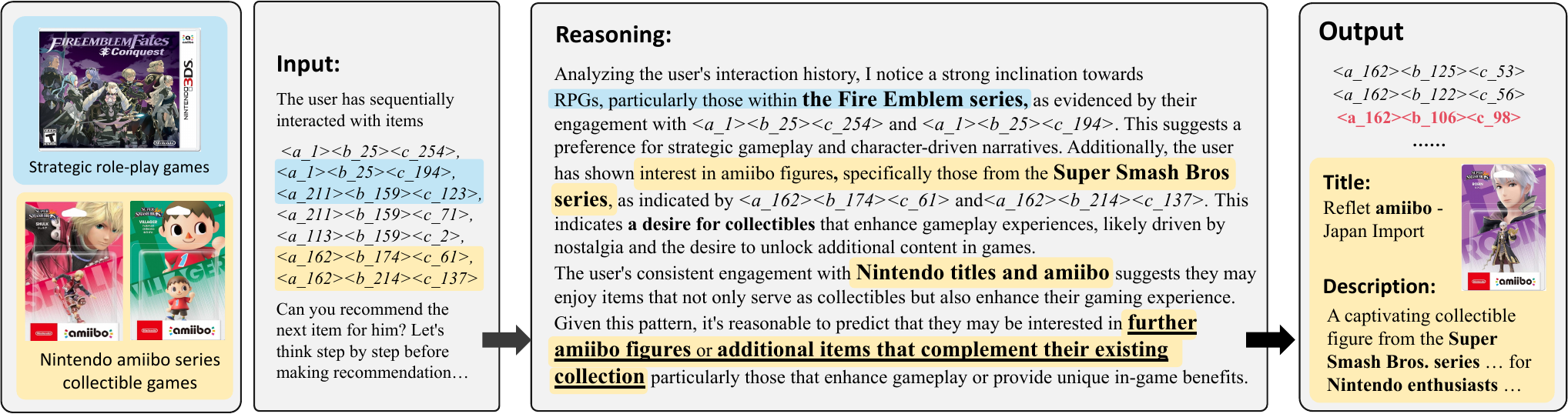}
    \vspace{-0.15in}
    \caption{Case study of how explicit reasoning yields effective and explainable recommendations.}
    \label{fig:case_study}
\end{figure*}

\subsubsection{Effect of Teacher Model on Corpus Enrichment.}
\ours benefits from the enriched pre-training corpus, which provides diverse and informative SID-language associations that strengthens SID-language alignment.
The quality and informativeness of the enriched corpus largely depend on how well the teacher model understands item semantics and enriches them with relevant world knowledge. To further investigate how different corpus construction strategies affect the final performance, we conduct a detailed ablation study where we compare corpus generated by GPT-4o-mini with three alternatives: (1) \textbf{Metadata}, where the enriched corpus is replaced by a rule-based concatenation of item metadata information without LLM-based rewriting; (2) \textbf{Qwen3-8B}, an open-source LLM with relatively limited capacity; and (3) \textbf{Qwen3-32B}, a stronger open-source LLM. We compare the performance of variant models trained with different enrichment corpora on the Games dataset and report the results by Recall@10 and NDCG@10.

As shown in Table~\ref{tab:ablation_teacher}, the model achieves the weakest performance when using rule-based metadata, indicating that directly concatenating metadata fields provides limited semantic expansion and insufficiently diverse SID-language contexts. Replacing metadata with Qwen3-8B brings marginal improvement, suggesting that a smaller teacher model may still be limited in producing rich and reliable semantic associations for SID alignment. In contrast, Qwen3-32B leads to a clear performance gain, showing that a stronger teacher can synthesize more informative corpora and thereby improve the effectiveness of SID-language alignment. GPT-4o-mini achieves the best performance among all variants, further confirming that the effectiveness of alignment is closely tied to the capability of the teacher model. Overall, stronger teachers tend to generate higher-quality and more semantically enriched corpora, which in turn helps the model acquire more effective reasoning ability for recommendation. The result also suggests that SIDReasoner naturally benefits from the continuous improvement of general LLMs, as stronger language models provide better semantic knowledge for constructing enriched pre-training corpora.

\begin{figure}[t]
    \centering
    \includegraphics[width = 0.98\linewidth]{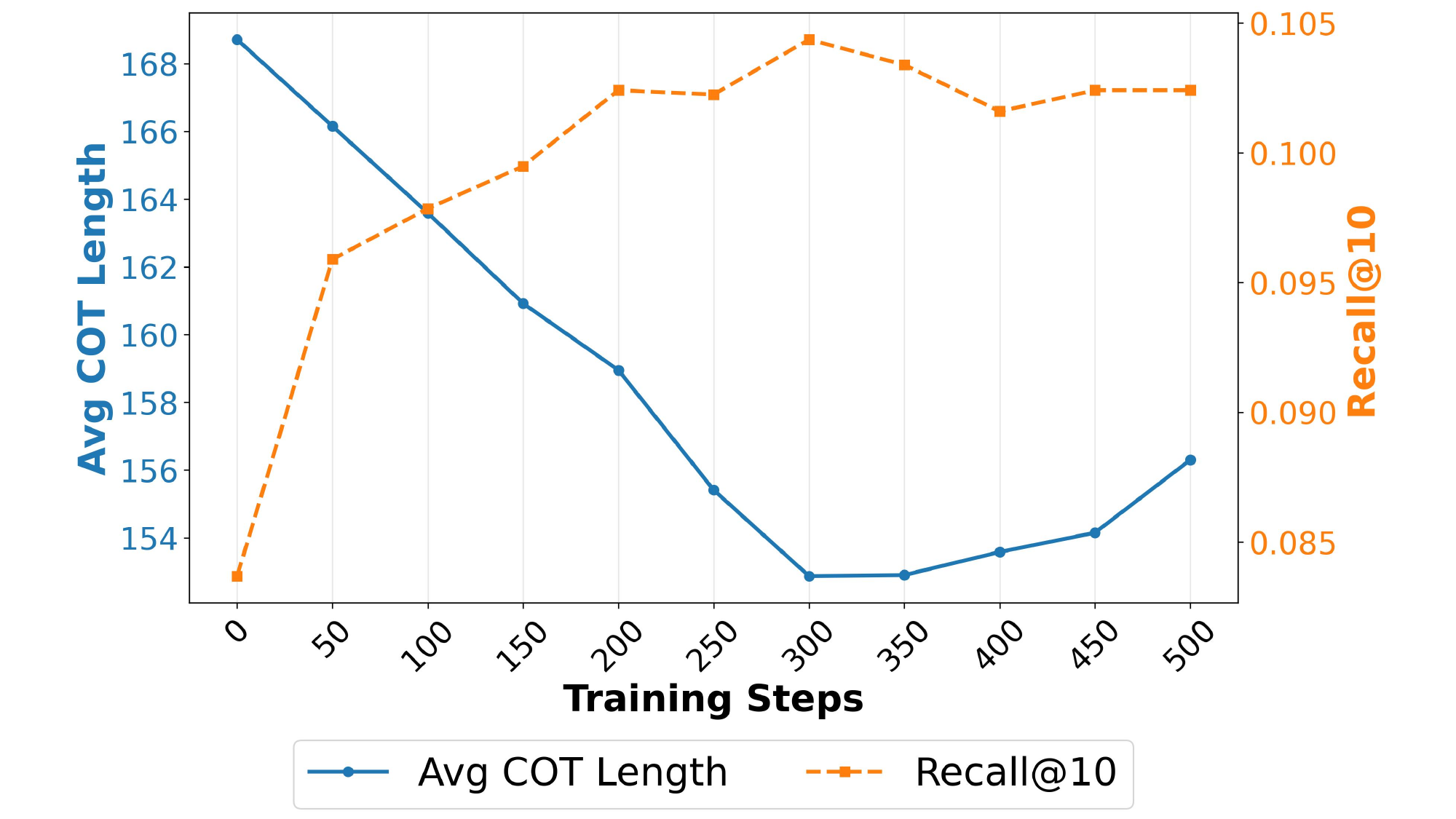}
    \vspace{-0.12in}
    \caption{Reasoning length and performance change during RL Training on Games Dataset.}
    \label{fig:model_study_change_of_cot}
    \vspace{-0.1in}
\end{figure}

\subsubsection{Evolution of Reasoning during RL Training.} 
Figure~\ref{fig:model_study_change_of_cot} illustrates the evolution of reasoning behavior during reinforcement learning on the Games dataset. Specifically, we track the average length of generated reasonings (blue curve) alongside the corresponding recommendation performance measured by Recall@10 (orange curve) at different training steps.
During RL training, the average reasoning length decreases notably in the early stages and converges to a shorter level, whereas recommendation performance steadily improves.
We hypothesize that this phenomenon stems from the nature of the reasoning patterns learned during SID-language alignment.
Before RL, the model primarily acquires its reasoning behaviors from samples generated by the teacher model (GPT-4o-mini).
While such teacher-generated reasoning provides a useful starting point, it is not necessarily optimized for recommendation and often contains redundant or uninformative reasoning patterns that do not contribute to prediction performance.
During RL, the model rapidly identifies and discards these ineffective reasoning components. As a result, the model progressively shifts toward shorter, more targeted reasoning trajectories that retain essential decision-relevant information while eliminating unnecessary intermediate steps. Ultimately, the model achieves better recommendation performance with more compact reasoning, indicating that effective recommendation reasoning does not require longer chains, but rather more efficient ones.

\subsubsection{Case Study}
\label{sec:case_study}

Figure~\ref{fig:case_study} presents a representative example of how explicit reasoning over SIDs enhances generative recommendation.
Given the user’s interaction history encoded as SID sequences, the model generates a reasoning trace that first summarizes the user’s key interests from historical interactions, including strategic role-playing games and Nintendo amiibo items.
Building on this summary, the model continues thinking and concludes that recommending additional Nintendo amiibo items is more appropriate given the user’s repeated engagement with such products.

Conditioned on this reasoning, the model prioritizes SIDs related to Nintendo amiibo, which eventually hits the target item successfully.
This example provides a transparent view into the model’s internal decision-making process, where intermediate reasoning explicitly reveals how user intent is inferred and subsequently used for recommendation.
Such reasoning directly shapes the SID decoding trajectory and the resulting recommendation list, rather than serving as a post-hoc explanation of the output.

\section{Conclusion}
In this work, we investigate how to enable effective reasoning in SID-based generative recommendation.
We propose \ours, a two-stage framework that effectively allows large language models to reason over itemic tokens.
Stronger SID–language alignment can help general reasoning abilities better generalize to recommendation tasks.
Motivated by this insight, our approach leverages multi-task alignment and teacher-assisted semantic expansion to establish strong semantic grounding between SID and language tokens.
Building on this foundation, outcome-driven reinforcement further guides the model to self-explore effective reasoning patterns in recommendation.
Extensive experiments on real-world datasets demonstrate that incorporating reasoning into generative recommendation yields competitive performance, together with improved cross-domain generalization and interpretability.

For future work, we plan to explore whether reasoning over SIDs yields further gains with larger backbone models, and to investigate the effectiveness of enriched SID–language alignment under substantially larger-scale training data. Due to limited time and computational resources, we leave scaling to larger models and datasets for future exploration.

\section{Acknowledgment}
This research is supported by National Natural Science Foundation of China (U25A20445).

\newpage
\bibliographystyle{ACM-Reference-Format}
\balance
\bibliography{references}

\clearpage
\appendix



\section{Baseline Methods} \label{appendix:baselines}
This section provides a brief introduction to the baselines used in our experiments.
\begin{itemize}[leftmargin=*]
\item \textbf{Caser} \cite{Caser} models sequential user behaviors by applying convolutional neural networks over user interaction sequences to capture local and global sequential patterns. It is trained to predict the next item based on the user’s historical interactions using the binary cross-entropy loss.

\item \textbf{GRU4Rec} \cite{GRU4Rec} utilizes GRU modules to capture sequential dependencies within user interaction sequences. It is trained to predict the next item in a user’s sequence based on previously purchased items. We use the binary cross-entropy loss as optimization objective during the training process.

\item \textbf{SASRec} \cite{SASRec} is a transformer-based recommender widely used in sequential recommendation. It leverages self-attention to capture long-range dependencies in user interaction sequences, enhancing the accuracy of future interaction predictions. Binary cross-entropy loss is used as the optimization objective.

\item \textbf{TIGER} \cite{TIGER} is a representative generative recommender that represents items as Semantic IDs and utilizes a Transformer model to autoregressively predict the Semantic IDs (SIDs) of next item. The model is optimized with cross-entropy loss.

\item \textbf{HSTU} \cite{HSTU} follows the formulation of TIGER, and models the recommendation task as sequence prediction over action and item tokens. Cross-entropy loss is used during model training.

\item \textbf{LETTER} \cite{LETTER} follows the structure of TIGER and further integrate collaborative signals into SIDs during the quantization stage. The model is optimized using the cross-entropy loss.

\item \textbf{LCRec} \cite{LCRec} adapts a pre-trained LLMs as the backbone of the generative recommendation. It further enhances SID understanding via a series of recommendation-related tasks. For fair comparison, we implement LCRec by with exactly same training inputs as our multi-task alignment.

\item \textbf{ReaRec} \cite{ReaRec} enhances the SASRec model with latent reasoning, and boosts recommendation effectiveness by introducing extra test-time compute. We use the Progressive Reasoning Learning method cross-Entropy loss.

\item \textbf{R\textsuperscript{2}ec} \cite{R2Rec} is constructed based on a pre-trained LLM. Given the historical interactions described in natural language, it leverages explicit textual reasoning before the recommendation decision. The model is trained using Proximal Policy Optimization (PPO) \cite{PPO} with recommendation-specific rewards.

\end{itemize}

\section{Alignment Task Formats} \label{appendix:alignment_formats}
In this section, we describe the instruction templates used for the alignment tasks, which are designed to bridge the gap between textual item information and their corresponding Semantic IDs. The eight templates fall into four functional categories, summarized below. For the complete ChatML-formatted system/user/assistant messages and concrete worked examples of every case, we refer the reader to our open-source repository.

\noindent \textbf{SID--Title Translation.} This category, comprising two templates in our open-source repository, establishes a bi-directional mapping between textual item titles and their corresponding SIDs. Given a title, the model is trained to emit the SID; given a SID, the model recovers the title. The mapping grounds the SID vocabulary in natural language, allowing the model to interchangeably reason about items in either representation.

\noindent \textbf{Generative Next-Item Prediction.} This category, comprising four templates in our open-source repository, requires the model to predict the next interaction by alternating input and output representations between item titles and SIDs, spanning all four directional variants (title$\rightarrow$title, title$\rightarrow$SID, SID$\rightarrow$title, and SID$\rightarrow$SID). Combining the four directions allows the model to capture sequential behavioral patterns while continuously reinforcing the correspondence between SIDs and textual descriptions.

\noindent \textbf{Alignment via Item-centric Semantic Enrichment.} This category, comprising one template in our open-source repository, asks the model to associate each SID with detailed, LLM-augmented item contexts. By interleaving the SID token throughout a coherent description that carries category, feature, and use-case information, the model binds the SID to a rich semantic context rather than to a sparse title alone.

\noindent \textbf{Sequential Recommendation via User-centric Reasoning Augmentation.} This category, comprising one template in our open-source repository, trains the model on hybrid narratives of interaction history that interleave natural language with Semantic IDs. By presenting the user's history as a coherent story---augmented with analyst-style reasoning over preference shifts---the model learns to understand user behavior and recommendation logic rather than merely pattern-matching on raw ID sequences.

\section{Prompts for Corpus Enrichment} \label{appendix:enrich_prompts}
We adopt a two-stage prompting framework to construct both the item-centric and user-centric enrichment corpora. We outline the role of each stage below; the verbatim prompt templates, output formatting constraints, and additional implementation details are released in our open-source repository.

\noindent \textbf{Item-centric Semantic Enrichment.}
\textbf{Stage 1 (Comprehensive Analysis)} prompts the model to reason over raw item metadata---title, brand, category, description, and features---and produce structured insights, including a refined description, primary use cases, target audiences, key feature highlights, and related keywords. This transforms sparse catalog entries into deep contextual knowledge. \textbf{Stage 2 (Integration)} then fuses all of the above information into a single coherent paragraph, under the explicit constraint that every mention of the item must use its SID token rather than its title. The resulting text serves as the supervision signal that binds an SID to a semantically dense description.

\noindent \textbf{User-centric Reasoning Enrichment.}
\textbf{Stage 1 (Reasoning Generation)} has the model adopt an analyst's persona and infer latent user preferences and behavioral shifts from the interaction history, producing a first-person reasoning monologue that expresses general interest directions without revealing the held-out next item. \textbf{Stage 2 (Narrative Integration)} merges the raw interaction sequence with the Stage~1 reasoning into a flowing natural-language paragraph in which every item is referenced exclusively by its SID. The output reads as a coherent story of the user's behavior, providing a denser training signal than the original ID sequence alone.

\end{document}